\documentstyle[preprint,aps]{revtex}

\begin{document}
\draft

\title{Superconductivity and spin fluctuations in the electron-doped infinitely-layered 
high $T_{c}$ superconductor Sr$_{0.9}$La$_{0.1}$CuO$_{2}$ ($T_{c}=$42 
K)}
\author{T. Imai$^{1,2*}$, and Charles P. Slichter$^{1,2,3}$}
\address{Department of Physics,$^1$ Science and Technology Center 
for Superconductivity,$^2$ \newline and Department of 
Chemistry$^3$,\linebreak
The University of Illinois at Urbana-Champaign
1110 West Green Street, Urbana, IL 61801-3080}
\author{Jonathan L. Cobb  
and  John T. Markert}
\address{Department of Physics, The University of Texas at Austin
Austin, TX78712-1081}
\maketitle


\vspace{2in}

$^{*}$Present Address:  Department of Physics, Massachusetts Institute of 
Technology,\newline
\#13-3149, Cambridge, MA 02139.

\vspace{0.5in}

Invited Paper to SNS-95 (Spectroscopies on Novel Superconductors, 
Stanford), \newline March, 1995. Published in J. Phys. and Chem. Solids {\bf 
56} 1921 (1995).\newline
Also presented at 1994 Aspen Winter Conference on Superconductivity 
(Invited),\newline  1994 M$^{2}$S-HTSC at Grenoble (Oral,
Physica {\bf C 235-240} (1994) 1627), \newline and 1994 APS meeting (Bull. 
Am. Phys. Soc. {\bf 39} (1994) 839.) 

\pagebreak

\begin{abstract}
We report $^{63}$Cu NMR studies on uniaxially aligned and unaligned powder samples of an electron-doped
infinitely-layered high $T_{c}$ cuprate superconductor Sr$_{0.9}$La$_{0.1}$CuO$_{2}$ ($T_{c}=$42 K).  
We measured the nuclear spin-lattice relaxation rate $1/T_{1}$ up to 800 K.  
We found from the temperature dependence of $1/T_{1}T$ that the wave-vector-averaged spin susceptibility 
is highly enhanced by spin-fluctuations in the normal state, resulting in a Curie-Weiss behavior.  
Below $T_{c}$, we did not observe a Hebel-Slichter coherence peak of $1/T_{1}$, suggesting an unconventional 
nature of the symmetry of the superconducting order parameter.  These results are quite similar to 
those observed for some hole-doped high $T_{c}$ cuprates.  
\end{abstract}

\pagebreak
\section{Introduction}
The mechanism of high temperature superconductivity in doped cuprate high $T_{c}$ superconductors 
is a very controversial issue.  So far, no consensus has been reached.  However various experiments 
have established that undoped parent compounds of high $T_{c}$ cuprates 
are spin $S=\frac{1}{2}$ antiferromagnets.  
Microscopic experimental probes including NMR (Nuclear Magnetic Resonance), 
NQR (Nuclear Quadupole Resonance) [1], neutron scattering[2], 
and Raman scattering[3] revealed that one can describe the magnetic 
properties of the undoped parent compounds based on the two dimensional
Heisenberg model reasonably well.  The next crucial step toward the complete 
understanding of high temperature superconductivity is to figure out the nature
of the carriers doped into the conducting CuO$_{2}$ planes, and the influence of 
mobile carriers on the electronic states of the undoped quasi 2d Heisenberg antiferromagnets.
The undoped compound La$_{2}$CuO$_{4}$ is a quasi 2d Heisenberg antiferromagnet with 
intraplane exchange interaction $J/k_{B}$ = 1520 K and bulk 3d Neel ordering 
temperature $T_{N}\sim$ 325 K [2].  Recently, after several years of painful 
trials and errors, the Illinois-Kyoto collaboration reported the first 
successful detection of $^{63,65}$Cu NQR in the paramagnetic state of 
La$_{2}$CuO$_{4}$[1].  
Their discovery of the NQR signal opened a way to investigate the influence of 
hole-doping in the entire doping range between the undoped La$_{2}$CuO$_{4}$ and the optimum 
hole-doped high $T_{c}$ superconductor La$_{1.85}$Sr$_{0.15}$CuO$_{4}$ ($T_{c}$=38 K) by directly observing 
the nuclear resonance in the CuO$_{2}$ planes.  They measured the temperature dependence 
of the nuclear spin-lattice relaxation rate $1/T_{1}$ and the Gaussian component of the 
spin-spin relaxation rate $1/T_{2G}$.  These quantities probe the imaginary and real 
parts of the weighted wave-vector {\bf q}-averaged dynamical electron spin susceptibility 
$\chi''({\bf q}, \omega=\omega_{n})$ and $\chi'({\bf q}, \omega=0)$, respectively, 
where $\omega$ is frequency and $\omega_{n}$ is the resonance frequency [4,5].  
Theoretical calculations for the 2d-Heisenberg model based on dynamical scaling[6], 
high temperature expansion[7], and finite cluster calculations[8] reproduce the experimental 
data of La$_{2}$CuO$_{4}$ quite well without any adjustable parameters.
The most surprising observation by Imai et al. is that $1/T_{1}$ levels off above 750 K 
at approximately the same value regardless of the doping level [1].  This finding 
demonstrates that at high temperatures the spin dynamics of the high $T_{c}$ superconductor 
La$_{1.85}$Sr$_{0.15}$CuO$_{4}$ retains essentially the same properties 
of the undoped antiferromagnet La$_{2}$CuO$_{4}$.  
In other words, in the first order approximation, copper d-spins in the superconducting phase may 
be viewed as localized moments at elevated temperatures [9].  More recently Sato and coworkers also 
reported that the Hall coefficient levels off at elevated temperatures 
for La$_{2-x}$Sr$_{x}$CuO$_{4}$ ($0.04\leq x\leq 0.15$) 
at approximately the same value [10].  This suggests that charge dynamics in the 
hole-doped superconducting phase may also keep the fundamental characteristic of 
the insulating phase at elevated temperatures.  On the other hand, King et al. 
reported that the dispersion relations of the bands crossing or close to 
the Fermi surface are very different in an electron-doped high $T_{c}$ superconductor 
Nd$_{1.85}$Ce$_{0.15}$CuO$_{4}$ compared with those in hole-doped materials [11].  
The temperature dependence of the penetration depth below $T_{c}$ in 
Nd$_{1.85}$Ce$_{0.15}$CuO$_{4}$ reported by Wu et al. is also regarded to be an 
evidence for the isotropic s-wave pairing realized in the electron-doped system [12], 
in contrast with the result for a hole-doped system 
YBa$_{2}$Cu$_{3}$O$_{6.95}$ by Hardy et al. 
which clearly indicates an unconventional nature, most likely to be d-wave pairing[13].
In view of these results, a naive question is what happens to the NMR properties 
if one dopes electrons instead of holes into the CuO$_{2}$ planes?  Unfortunately it 
is not easy to answer this question because of the following two reasons.  
Firstly, the physical properties of the well-known electron-doped high $T_{c}$ superconductor 
Nd$_{1.85}$Ce$_{0.15}$CuO$_{4}$ are extremely sensitive to the oxygen 
content as may be seen in the results of Kambe et al.[14].  
This makes preparation of high quality samples as well as 
reproducible studies very hard.  Secondly, the presence of 
the large Nd moments masks the intrinsic physical properties of CuO$_{2}$ planes.  
In particular, $^{63}$Cu NMR experiments on Nd$_{1.85}$Ce$_{0.15}$CuO$_{4}$ are almost completely 
dominated by the large Nd moments as demonstrated by Zheng et al.[15].  
In this paper, we discuss our recent NMR results [16] 
for the electron-doped high $T_{c}$ superconductor Sr$_{0.9}$La$_{0.1}$CuO$_{2}$ ($T_{c}$ = 42 K) [17], 
and compare the results with those of hole-doped materials.  
This material is suited for NMR experiments because unlike the case 
of Nd$_{1.85}$Ce$_{0.15}$CuO$_{4}$ there are no unpaired spins outside 
the CuO$_{2}$ planes 
in Sr$_{0.9}$La$_{0.1}$CuO$_{2}$ [18].  This enabled us to obtain information intrinsic 
to the electron-doped CuO$_{2}$ planes from $^{63}$Cu NMR for the first time.  
In addition, the infinitely-layered structure of Sr$_{0.9}$La$_{0.1}$CuO$_{2}$ allows us 
to investigate the possible influence of interlayer coupling between 
adjacent CuO$_{2}$ planes on the physical properties of high $T_{c}$ cuprates [19].  
Our key finding is that the spin fluctuation and superconducting properties 
of Sr$_{0.9}$La$_{0.1}$CuO$_{2}$ probed by NMR exhibited very similar results to those of hole-doped compounds.

\section{Experimental}
Polycrystalline samples of Sr$_{0.9}$La$_{0.1}$CuO$_{2}$  were prepared by utilizing high pressure apparatus at Texas.  
Details of the preparation of the samples and their characterization are discussed elsewhere [17].  
The hard pellet samples were crushed and sieved to obtain fine ceramic powder samples.  
A part of the fine powder was uniaxially aligned in external magnetic field of 8.3 Tesla in Stycast 1266 
resin at room temperature.  Since Stycast reduces the high $T_{c}$ cuprate samples above approximately 500 K, 
NMR measurements above 400 K were carried out utilizing a powder sample with no epoxy sealed in a nitrogen gas atmosphere.  
After we completed the measurements of the NMR line shape and $1/T_{1}$ up to 600 K, we cooled down the sample 
to room temperature to check the reproducibility.  We found that $1/T_{1}$ at room temperature was identical 
to that before the heat cycling.  We repeated the same procedures after the measurements at 800 K, 
again confirming the reproducibility of $1/T_{1}$ at room temperature, 
though we found a weak $^{63}$Cu NMR 
signal from small amount of copper metal reduced from the sample.
All the NMR measurements were carried out at Illinois using standard home made pulsed NMR spectrometers.  
Superconducting magnets operating either at 8.3 Tesla or at 9.4 Tesla were utilized.  
NMR line shape was measured by integrating the spin echo signal point by point scanning the resonance frequency.  
$1/T_{1}$ was measured by inverting the population of nuclear magnetization with a 180 degree pulse.

\section{Results and Discussions}
\subsection{Static Properties}
In Fig.1, we present the $^{63}$Cu and $^{65}$Cu NMR line shape of an unaligned powder sample of 
Sr$_{0.9}$La$_{0.1}$CuO$_{2}$ measured at 295 K.  There is a small ÒshoulderÓ in the broad line shape 
at the higher frequency side of each of the $^{63}$Cu and $^{65}$Cu NMR spectra, characteristic 
of the NMR line shape with anisotropic Knight shift with axial symmetry. In fact we 
confirmed that the angular dependence of the peak position of $^{63}$Cu NMR in the aligned 
powder sample fits the theoretical curve expected for the case of an anisotropic Knight shift [20] quite well.
In Fig. 2 (a) and (b), we present the $^{63}$Cu NMR line shapes observed for the uniaxially 
aligned powder sample with the magnetic field applied parallel to the crystal c-axis 
and the ab-plane, respectively.  One notices that there is a relatively sharp transition 
and broad tails for both higher and lower frequency sides.  We attribute the sharp peak 
to the transition from nuclear spin quantum number $I_{z}=1/2$ to 
$I_{z}=-1/2$ (i.e. the central transition), 
and the broad tail to the transition from $I_{z}=3/2$ to $1/2$ and from 
$I_{z}=-3/2$ to $-1/2$ (i.e. satellite transitions).  
We verified the assignment by comparing the nutation curves of the spin echo signal 
as a function of $t_{w}$, the width in time of the RF exciting pulse for Cu metal, 
the $I_{z}=1/2$ to $I_{z}=-1/2$ transition of the planar Cu site of 
YBa$_{2}$Cu$_{3}$O$_{7}$, and the central 
peak of the aligned powder of Sr$_{0.9}$La$_{0.1}$CuO$_{2}$ under identical experimental conditions.  
For Cu metal, a 90 degree pulse occurs when $\gamma H_{1}t_{w}=\pi/2$, 
where $\gamma$ and $H_{1}$ are the nuclear 
gyromagnetic ratio and the strength of RF magnetic field, respectively.  
For the $+1/2$ to $-1/2$ transition, a 90 degree pulse occurs when 
$2\gamma H_{1}t_{w}=\pi/2$ 
due to the matrix element effects [20].  We found that the results for 
YBa$_{2}$Cu$_{3}$O$_{7}$ and Sr$_{0.9}$La$_{0.1}$CuO$_{2}$ were identical, as shown in the inset to Fig.1, 
while $t_{w}$ for YBa$_{2}$Cu$_{3}$O$_{7}$ and Sr$_{0.9}$La$_{0.1}$CuO$_{2}$ was 2 times shorter than that for Cu metal.  
From the separation of the satellite transitions, 
we estimate the nuclear quadrupole coupling $\nu_{Q}$ to be 4 MHz or below.  
In general the peak position of the NMR line for the $I_{z}=1/2$ to $I_{z}=-1/2$ 
transition is shifted by both the NMR Knight shift, $^{63}K$, and the second 
or higher order effects due to the quadrupole interaction [20].  
We confirmed that the relative shift of the apparent peak position is 
identical for both $^{63}$Cu and $^{65}$Cu isotopes for 8.3 and 9.4 Tesla 
when the magnetic field is applied parallel to the aligned c-axis.  
This indicates that the main principle axis of the quadrupole coupling 
tensor is parallel to the c-axis, and the contribution from the quadrupolar effects to the NMR 
shift is null in the case.   
By taking $\nu_{Q}=$4 MHz, the contribution of the quadrupole effects to the shift for the field 
applied parallel to the ab-plane is 0.03\% at 8.3 Tesla.  
By taking into account these considerations, our preliminary results 
on the temperature dependence of Knight shifts are as follows.  
$^{63}K_{c}$ saturates at elevated temperatures and decreases below room temperature 
from 0.92\% at 295 K to 0.86\% at $T_{c}$=42K.  In contrast, 
$^{63}K_{ab}$ = 0.29 \% is independent of temperature.  
These results are in remarkable contrast with the case of underdoped region of the hole doped systems, 
where $^{63}K_{c}$ is temperature independent and $^{63}K_{ab}$ decreases with temperature [21].  
The hyperfine coupling tensor in Sr$_{0.9}$La$_{0.1}$CuO$_{2}$ seems to 
be rather different from that in YBa$_{2}$Cu$_{3}$O$_{x}$ and 
La$_{2-x}$Sr$_{x}$CuO$_{4}$.

\subsection{Spin Dynamics}
In Fig. 3 we present an example of the dependence of the echo integral
$M(t)$ on the delay time $t$ of the echo sequence after a 180 degree inversion pulse.  
The result fits very well the solution of the rate equation for the 
$I_{z}=1/2$ to $I_{z}=-1/2$ transition 
of nuclear spin $I=3/2$, $M(t) = A - B [0.9 exp(-6t/T_{1}) + 0.1 
exp(-t/T_{1})]$, where A and B are constants, 
and $T_{1}$ is the spin-lattice relaxation time.  
The temperature dependence of the spin-lattice relaxation rate $1/T_{1}$ measured for an aligned powder 
sample under the external magnetic field 8.3 Tesla applied parallel 
to the crystal c-axis or ab-plane is presented in Fig.4.  
The anisotropy of $1/T_{1}$, $^{63}R$, is approximately $^{63}R\sim$2.6.  
This is comparable to but somewhat smaller than the value observed for 
undoped La$_{2}$CuO$_{4}$ [1] 
and hole-doped YBa$_{2}$Cu$_{3}$O$_{7}$ [22], $^{63}R\sim$3.7, and much smaller than that 
for an antiferromagnet CuO, $^{63}R\sim$8 [23].  
The large value of $^{63}R$ in CuO is due to the absence of the transferred hyperfine interaction.  
Our observation suggests that the mechanism of transferred hyperfine interaction between nearest 
neighbor Cu sites [24] are present in electron-doped systems, too.  
Mild scattering of the data for $(1/T_{1})_{c}$ caused by the broad line width 
and the resulting contamination by the satellite transition make it difficult 
to obtain the precise value and temperature dependence of $^{63}R$.  
It appears that $^{63}R$ does not show any strong temperature dependence above $T_{c}$.
Above 450 K, we found that the ceramic powder sample was reduced and destroyed in stycast 1266.  
To avoid this, we carried out measurements of $1/T_{1}$ for an unaligned powder sample 
at the peak position of the powder spectrum at higher temperatures. We normalized the data for powder, 
$(1/T_{1})_{powder}$, to that measured for aligned powder, 
$(1/T_{1})_{ab}$, at 295 K by multiplying with a factor $1.4$.  
We plotted the data of $(1/T_{1})_{ab}$ and $1.4(1/T_{1})_{powder}$  together in Fig. 5.  Since the normalized 
results showed good agreement also at 77 K, the normalization procedure should not alter any essential 
aspect of the data.
It is clear from Fig. 5 that the results for the electron-doped compound is 
semi-quantitatively similar to that of various hole-doped compounds.  
A high relaxation rate $1/T_{1}$ with negative curvature in its temperature 
dependence indicates that the electronic states can not be described by the canonical Fermi liquid state, 
and strong correlation effects have to be taken into account.  As shown in Fig. 5(c), 
the temperature dependence of $T_{1}T$ follows a Curie-Weiss law, 
$T_{1}T= c(T-T^{*})$, 
where $c=3\times10^{-4}$ sec and $T^{*} = -174$ K are constants, in the entire temperature range 
between $T_{c}$ = 42 K and 800 K.  As demonstrated by Moriya [4], 
$1/T_{1}$ probes the weighted {\bf q}-average 
of the imaginary part of the electron spin susceptibility.  
Therefore our finding indicates that the wave vector {\bf q} averaged spin susceptibility 
in the electron doped superconductor Sr$_{0.9}$La$_{0.1}$CuO$_{2}$ satisfies a Curie-Weiss law.  
The negative value of $T^{*}$ means that the electron spin system in the electron-doped 
CuO$_{2}$ planes are not approaching the Neel ordered ground state.  
These findings are essentially the same as those previously reported for YBa$_{2}$Cu$_{3}$O$_{7}$ 
by Barrett et al. [25]  
We also attempted to fit $1/T_{1}$ to a power law, $1/T_{1} \sim T^{n}$ 
($n\sim 1/2$), 
as suggested by Ren and Anderson for a Luttinger liquid system [26].  
Though our data is in general accord with a power-law behavior with $n < 
1$, we found that the fit is not good.
One unique feature of Sr$_{0.9}$La$_{0.1}$CuO$_{2}$ is in the strong interlayer coupling.  
In fact the undoped parent compound (Ca$_{0.85}$Sr$_{0.15}$)CuO$_{2}$ has the highest Neel ordering temperature, 540K, 
among various cuprates [27].  Moreover Cobb and Markert found that the superconducting anisotropy 
is relatively small in Sr$_{0.9}$La$_{0.1}$CuO$_{2}$.  Their observation is an evidence 
for strong interlayer coupling in the electron doped compound [28].    
The interlayer coupling of CuO$_{2}$ bi-layers has been proposed to be responsible for the spin gap behavior 
in underdoped YBa$_{2}$Cu$_{3}$O$_{x}$ and YBa$_{2}$Cu$_{4}$O$_{8}$, 
where $1/T_{1}$T shows a significant reduction above $T_{c}$ [21].  
Though one may anticipate that the spin gap behavior is even enhanced in the infinitely-layered 
compound compared with the bi-layer systems, we did not observe spin gap 
behavior in Sr$_{0.9}$La$_{0.1}$CuO$_{2}$.  
As presented in Fig. 5(b), $1/T_{1}T$ increases monotonically down to $T_{c}$.  
The interplay between the doping level, 
the nature of doping (hole or electron), interlayer coupling, and in-plane anisotropy [29] needs 
further investigation to clarify the mechanism of spin gap behavior observed in some hole-doped systems.
Another important finding in our data is that $1/T_{1}$ exhibits monotonic and sudden 
decrease just below $T_{c}$.  
Various model calculations showed that isotropic s-wave superconducting 
pairing results in a Hebel-Slichter coherence peak [30] just 
below $T_{c}$ even for the systems with strong antiferromagnetic spin fluctuations.  
The absence of the coherence peak in the magnetic field of 8.3 Tesla 
makes the s-wave scenario unlikely in the electron-doped superconductor Sr$_{0.9}$La$_{0.1}$CuO$_{2}$.

\section{Summary}
We reported our $^{63}$Cu NMR measurements for an electron-doped infinitely-layered high 
$T_{c}$ superconductor Sr$_{0.9}$La$_{0.1}$CuO$_{2}$.  This material is unique in the sense that the doping 
is made by adding electrons into CuO$_{2}$ planes instead of holes, and the interlayer coupling 
might be strongest among various cuprates.  Our experimental results indicate that the 
spin dynamics in the electron-doped Sr$_{0.9}$La$_{0.1}$CuO$_{2}$ show essentially the same temperature dependence 
as those in hole-doped compounds.  The symmetry of the orbital pairing in the superconducting state 
is unlikely to be isotropic s-wave, again analogous to the case of hole-doped systems.  
Investigation of further details of some aspects of our experiments including NMR shifts, 
hyperfine coupling, and the comparison with the undoped insulating antiferromagnet 
Ca$_{0.85}$Sr$_{0.15}$CuO$_{2}$ are under way.

\section{Acknowledgment}
The work at Illinois was supported by NSF-DMR 91-20000 
through the Science and Technology Center for Superconductivity, and b
y DOE-DEFG-02-9ER45439 through the Frederick Seitz Materials Research Laboratory.  
The work at Texas was supported by NSF-DMR-9158089, and by the Robert Welch Foundation under F-1191. 
 
\pagebreak 
\section{References}
[1] T. Imai, C. P. Slichter, K .Yoshimura, and K. Kosuge, Phys. Rev. 
Lett. {\bf 70} (1993) 1002. T. Imai, C. P. Slichter, 
K .Yoshimura, M. Katoh, and K. Kosuge, Phys. Rev. Lett. {\bf 71} (1993) 1254. T. Imai, C. P. Slichter, K. Yoshimura, M. Katoh and K. Kosuge,
Physica {\bf B197}, 601 (1994).\newline
[2] K. Yamada, K. Kakurai, Y. Endoh, T. R. Thurston M. A. Kastner, R. J. Birgeneau, G. Shirane, Y. Hidaka, 
and T. Murakami, Phys. Rev. {\bf B40} (1989) 4557. S. M. Hayden, G. Aeppli, 
H. A. Mook, S-W. Cheong, and Z. Fisk, Phys. Rev. {\bf B42} (1990) 10220.\newline
[3] For example, see S. Sugai and Y. Hidaka, Phys. Rev. {\bf B40} 
(1991)809, I. Tomeno et al., Phys. Rev. {\bf B43}(1991)3009.\newline
[4] T. Moriya, Prog. Theor. Phys. (Kyoto), {\bf 16} (1956) 33.\newline
[5] C. H .Pennington and C. P. Slichter, Phs. Rev. Lett. {\bf 66} (1991) 381.\newline
[6] S. Chakravarty, B. I. Halperine, and D. R. Nelson, Phys. Rev. {\bf B39}, (1989) 2344.   
S. Chakravarty and R. Orbach, Phys. Rev. Lett. {\bf 64} (1990) 1254.   
A. Chubukov and S. Sachdev, Phys. Rev. Lett. {\bf 71} (1993) 169. 
A. Chubukov, S. Sachdev and A. Sokol, Phys. Rev. {\bf B49} (1994) 9052.\newline
[7] R.P.P. Singh and M. P. Gelfand, Phys. Rev. {\bf B42} (1990) 996.  
A. Sokol, R. L. Glenister, and R. P. P. Singh, Phys. Rev. Lett. {\bf 72} (1993) 1549.\newline
[8] A. Sokol, E. Gagliano, and S. Bacci, Phys. Rev. {\bf B47} (1993) 14646.\newline
[9] Q. Si, J. H. Kim, J. P. Lu, and K. Levin, Phys. Rev. {\bf B42} (1990) 1033.\newline
[10] T. Nishikawa, J. Takeda, and M. Satoh, J. Phys. Soc. Jpn. {\bf 63} (1994)1441.\newline
[11] D.M. King, Z.X. Shen, D.S. Dessau, B.O. Wells, W.E. Spicer, 
A.J. Arko, D.S. Marshall, J. DiCarlo, A.G. Loeser, C.H. Park, E.R. Ratner, J.L. Peng, Z.Y. Li, and R.L. Greenn, 
Phys. Rev. Lett. {\bf 70}(1993)3159.\newline
[12] D.H. Wu, J. Mao, S. N. Mao, J. L. Peng, X. X. Xi, T. Venkatesan, 
R. L. Greene, and S. M. Anlage, Phys. Rev. Lett. {\bf 70}(1992)85.\newline
[13] W. N. Hardy, D. A. Bonn, D. C. Morgan, R. Liang and K. Zhang, Phys. 
Rev. Lett. {\bf 70}(1993)3999.\newline
[14] S. Kambe, H. Yasuoka, H. Takagi, S. Uchida, and Y. Tokura, J. Phys. 
Soc. Jpn., {\bf 60} (1991) 400.\newline
[15] G. Zheng, Y. Kitaoka, Y. Oda,  and K. Asayama, J. Phys. Soc. 
Jpn. {\bf 58} (1989) 1910.\newline
[16] T. Imai, C. P. Slichter, K. Yoshimura, M. Katoh, K. Kosuge, J. L. 
Cobb and J. T. Markert, Physica {\bf C235-240} (1994) 1627.\newline
[17] M. G. Smith, A. Manthiram, J. Zhou, J. B. Goodenough, and J. T. Markert, 
Nature {\bf 351} (1991) 540.
J. L. Cobb, A. Morosoff, L. Stuk and J. T. Markert, Physica {\bf 
B194-196}(1994)2247.\newline
[18] We found that the nuclear spin-lattice relaxation rate of 
$^{139}$La is very slow in Sr$_{0.9}$La$_{0.1}$CuO$_{2}$.  
This indicates that the valence of La is three with no unpaired spin in the 5d-orbital.\newline
[19] A. J. Millis and H. Monien, Phys. Rev. Lett. {\bf 70} (1993) 2810.  
M. Ubbens and P. A. Lee, Phys. Rev. {\bf B50} (1994) 438.\newline
[20] C. P. Slichter, Principles of Magnetic Resonance, 3rd ed. (Springer-Verlag, Heidelberg, 1990).\newline
[21] M. Taligawa, A.P. Reyes, P.C. Hammel, J. D. Thompson, R.H. Hefner, 
Z. Fisk, and K.C. Ott, Phys. Rev. {\bf B43}(1991)247.\newline
[22] C. H. Pennington, D. J. Durand, C. P. Slichter, J. P. Rice, E. D. Bukowski, 
and D. M. Ginsberg, Phys. Rev. {\bf B39} (1988) 2902.\newline
[23] T. Imai, Thesis, The University of Tokyo, 1991.\newline
[24] F. Mila and T. M. Rice, Physica {\bf C157} (1989) 561.\newline
[25] S. E. Barrett, D. J. Durand, C. H. Pennington, C. P. Slichter, 
T. A. Friedmann, J. P. Rice, and D. M. Ginsberg, Phys. Rev. {\bf B41} (1990) 6283.\newline
[26] P. W. Anderson and Y. Ren, in The Proceedings of Los Alamos symposium on 
High Temperature Superconductivity, eds. K. S. Bedell, D. Coffey, D. E. Meltzer, 
D. Pines and J. R. Schrieffer, (Addison and Wesley, California, 1990).\newline
[27] A. Keren, L. P. Le, G. M. Luke, B. J. Sternlieb, W. D. Wu, Y. J. Uemura, S. Tajima, and S. Uchida, 
Phys. Rev. {\bf B48}, 12926 (1993).\newline
[28] J. L. Cobb and J. T. Markert, Physica {\bf C226} (1994) 235.\newline
[29] H.-Q. Ding, Phys. Rev. Lett. {\bf 68} (1992)1927.\newline
[30] L.C. Hebel and C.P.Slichter, Phys. Rev. {\bf 113}(1959)1504.\newline

\pagebreak

\section{Figure Captions}

Figure 1
	
$^{63}$Cu and $^{65}$Cu NMR spectra observed for the unaligned powder sample at 295 K in 8.3 Tesla.  
Inset:  Echo intensity of the central peak of 
Sr$_{0.9}$La$_{0.1}$CuO$_{2}$ and YBa$_{2}$Cu$_{3}$O$_{7}$ observed for the aligned 
powder sample plotted as a function of the width in time of the pulse width.

Figure 2	

$^{63}$Cu NMR spectra observed at 77 K for the uniaxially aligned powder sample of Sr$_{0.9}$La$_{0.1}$CuO$_{2}$. 
The magnetic field of 8.3 Tesla is applied parallel to (a) aligned c-axis, and (b) aligned ab-plane.

Figure 3	

Plot of the normalized intensity of echo integral $M(t)$ at time t after inversion 
of the population of spins.  Mo is the saturated value of the echo intensity 
for infinitely large $t$.  
Solid line is the fit to the solution of the rate equation.

Figure 4
	
Temperature dependencies of $^{63}(1/T_{1})_{ab,c}$ measured for the uniaxially aligned powder sample.

Figure 5
	
(a)Temperature dependence of $^{63}(1/T_{1})_{ab}$ obtained from the measurement for aligned powder 
sample (filled circles) and for unaligned powder sample (unfilled circles, 
normalized by multiplying a factor $1.43$).  
(b) $^{63}(1/T_{1}T)_{ab}$.  (c) $^{63}(T_{1}T)_{ab}$.  Solid line is a fit to 
Curie-Weiss law, $^{63}(T_{1}T)_{ab} = 3\times 10^{-4}(T+174)$.

\end{document}